\documentclass[sigconf]{acmart}
\usepackage{graphicx}
\usepackage{amsmath}
\usepackage{amsthm}
\usepackage{amsfonts}
\usepackage{multirow}
\usepackage{multicol}
\usepackage{multirow}
\usepackage{booktabs}
\usepackage{makecell}
\usepackage{color}
\usepackage{utfsym}
\usepackage{bm}
\usepackage{subfig}
\usepackage{float}
\AtBeginDocument{%
  }

\setcopyright{rightsretained}
\copyrightyear{2026}
\acmYear{2026}
\setcopyright{cc}
\setcctype{by}
\acmConference[WWW '26] {Proceedings of the ACM Web Conference 2026}{April 13--17, 2026}{Dubai, United Arab Emirates.}
\acmBooktitle{Proceedings of the ACM Web Conference 2026 (WWW '26), April 13--17, 2026, Dubai, United Arab Emirates}
\acmPrice{}
\acmDOI{10.1145/3774904.3792311}
\acmISBN{979-8-4007-2307-0/2026/04}

\makeatletter
\gdef\@copyrightpermission{
 \begin{minipage}{0.3\columnwidth}
  \href{https://creativecommons.org/licenses/by/4.0/}{\includegraphics[width=0.90\textwidth]{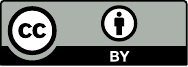}}
 \end{minipage}\hfill
 \begin{minipage}{0.7\columnwidth}
  \href{https://creativecommons.org/licenses/by/4.0/}{This work is licensed under a Creative Commons Attribution International 4.0 License.}
 \end{minipage}
 \vspace{5pt}
}
\makeatother

\acmSubmissionID{1512}

\begin{document}

%%
%% The "title" command has an optional parameter,
%% allowing the author to define a "short title" to be used in page headers.
\title[Hyperbolic Multimodal Generative Representation Learning for\\Generalized Zero-Shot Multimodal Information Extraction]{
    Hyperbolic Multimodal Generative Representation Learning for Generalized Zero-Shot Multimodal Information Extraction
}

%%
%% The "author" command and its associated commands are used to define
%% the authors and their affiliations.
%% Of note is the shared affiliation of the first two authors, and the
%% "authornote" and "authornotemark" commands
%% used to denote shared contribution to the research.
%%
%% The "author" command and its associated commands are used to define
%% the authors and their affiliations.
%% Of note is the shared affiliation of the first two authors, and the
%% "authornote" and "authornotemark" commands
%% used to denote shared contribution to the research.
\author{Baohang Zhou}
\affiliation{%
\institution{School of Software\\Tiangong University}
\city{Tianjin}
\country{China}
}
\email{zhoubaohang@tiangong.edu.cn}

\author{Kehui Song}
\authornote{Corresponding author.}
\affiliation{%
\institution{School of Software\\Tiangong University}
\city{Tianjin}
\country{China}
}
\email{songkehui@tiangong.edu.cn}

\author{Rize Jin}
\affiliation{%
\institution{School of Software\\Tiangong University}
\city{Tianjin}
\country{China}
}
\email{jinrize@tiangong.edu.cn}

\author{Yu Zhao}
\affiliation{%
\institution{College of Computer Science\\VCIP, DISSec, TMCC, TBI Center\\Nankai University}
\city{Tianjin}
\country{China}
}
\email{zhaoyu@dbis.nankai.edu.cn}

\author{Xuhui Sui}
\affiliation{%
\institution{College of Computer Science\\VCIP, DISSec, TMCC, TBI Center\\Nankai University}
\city{Tianjin}
\country{China}
}
\email{suixuhui@dbis.nankai.edu.cn}

\author{Xinying Qian}
\affiliation{%
\institution{College of Computer Science\\VCIP, DISSec, TMCC, TBI Center\\Nankai University}
\city{Tianjin}
\country{China}
}
\email{qianxinying@dbis.nankai.edu.cn}

\author{Xingyue Guo}
\affiliation{%
\institution{College of Computer Science\\VCIP, DISSec, TMCC, TBI Center\\Nankai University}
\city{Tianjin}
\country{China}
}
\email{guoxingyue@dbis.nankai.edu.cn}

\author{Ying Zhang}
\affiliation{%
\institution{College of Computer Science\\VCIP, DISSec, TMCC, TBI Center\\Nankai University}
\city{Tianjin}
\country{China}
}
\email{yingzhang@nankai.edu.cn}

%%
%% The abstract is a short summary of the work to be presented in the
%% article.
\begin{abstract}
    Multimodal information extraction (MIE) constitutes a set of essential tasks aimed at extracting structural information from Web texts with integrating images, to facilitate the structural construction of Web-based semantic knowledge.
    To address the expanding category set including newly emerging entity types or relations on websites, prior research proposed the zero-shot MIE (ZS-MIE) task which aims to extract unseen structural knowledge with textual and visual modalities.
    However, the ZS-MIE models are limited to recognizing the samples that fall within the unseen category set, and they struggle to deal with real-world scenarios that encompass both seen and unseen categories.
    The shortcomings of existing methods can be ascribed to two main aspects. On one hand, these methods construct representations of samples and categories within Euclidean space, failing to capture the hierarchical semantic relationships between the two modalities within a sample and their corresponding category prototypes. On the other hand, there is a notable gap in the distribution of semantic similarity between seen and unseen category sets, which impacts the generative capability of the ZS-MIE models.
    To overcome the above disadvantages, we delve into the generalized zero-shot MIE (GZS-MIE) task and propose the hyperbolic multimodal generative representation learning framework (HMGRL).
    The variational information bottleneck and autoencoder networks are reconstructed with hyperbolic space for modeling the multi-level hierarchical semantic correlations among samples and prototypes.
    Furthermore, the proposed model is trained with the unseen samples generated by the decoder, and we introduce the semantic similarity distribution alignment loss to enhance the model's generalization performance.
    Experimental evaluations on two benchmark datasets underscore the superiority of HMGRL compared to existing baseline methods.
\end{abstract}

%%
%% The code below is generated by the tool at http://dl.acm.org/ccs.cfm.
%% Please copy and paste the code instead of the example below.
%%


%%
%% Keywords. The author(s) should pick words that accurately describe
%% the work being presented. Separate the keywords with commas.

% \received{20 February 2007}
% \received[revised]{12 March 2009}
% \received[accepted]{5 June 2009}

%%
%% By default, the full list of authors will be used in the page
%% headers. Often, this list is too long, and will overlap
%% other information printed in the page headers. This command allows
%% the author to define a more concise list
%% of authors' names for this purpose.
\renewcommand{\shortauthors}{Baohang Zhou et al.}
\newcommand{\figref}[1]{Figure \ref{#1}}
\newcommand{\tbref}[1]{Table \ref{#1}}
\renewcommand{\eqref}[1]{Eqn. \ref{#1}}

%%
%% This command processes the author and affiliation and title
%% information and builds the first part of the formatted document.
\maketitle

\section{Introduction}
Multimodal information extraction (MIE)~\cite{9961954} encompasses a set of core tasks aimed at extracting structured information from texts in combination with images to build a multimodal knowledge graph~\cite{DBLP:journals/corr/abs-2507-20738,DBLP:conf/sigir/ZhaoZZQSC24}.
\begin{figure}[t]
    \centering
    \includegraphics[width=\columnwidth]{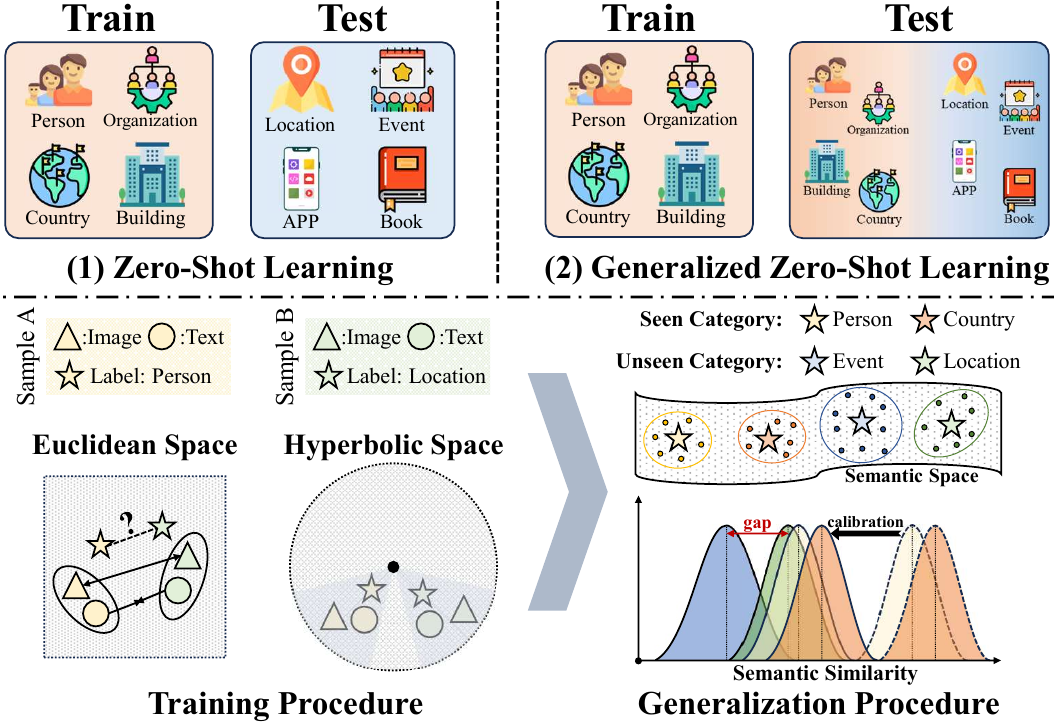}
    \caption{
        The upper part shows the comparison of the zero-shot learning and generalized zero-shot learning.
        The lower part represents the key challenges of the generalized zero-shot multimodal information extraction.
    }
    \label{fig:introduction}
    \vspace{-9pt}
\end{figure}
The MIE tasks are specifically crafted to achieve distinct goals, such as multimodal named entity typing (MET)~\cite{DBLP:journals/nn/ZhouZSSZY25} and multimodal relation extraction (MRE)~\cite{DBLP:conf/mm/ZhengFFCL021}.
Compared to information extraction (IE) models, multimodal methods are exploited to utilize the interconnections between textual and visual information using various fusion strategies, ultimately boosting performance on different tasks~\cite{zhou-etal-2024-mcil}.
To address the challenge posed by the continuously emerging new types of entities or relations, incorporating zero-shot learning into text-based information extraction (ZS-IE) models has enabled the recognition of unseen entity types or relations without the need for further training.
Current ZS-IE methods focus on the textual modality and extract entity features for building representations of type or relation prototypes. 
The names of types or relations are regarded as prototypical knowledge for bridging semantic correlation between samples and categories~\cite{DBLP:conf/coling/MaCG16,chen-li-2021-zs}.
To identify novel categories implied in multimodal data, \citet{DBLP:conf/www/ZhouZZSY25} firstly proposed the zero-shot MIE (ZS-MIE) task and introduced the specific architecture and training strategy.
Previous studies are grounded in the zero-shot learning~\cite{DBLP:journals/pami/PourpanahALZWLWW23} where the category sets of the training set and the test set are disjoint.

The current ZS-IE and ZS-MIE models are confined to identifying samples solely within the unseen category set.
Zhou et al. incorporated zero-shot learning into MET and MRE tasks, enabling the recognition exclusively of unseen entity types or relations during the testing phase~\cite{DBLP:journals/nn/ZhouZSSZY25,DBLP:conf/www/ZhouZZSY25}.
In practical situations, nevertheless, the contents on social media encompass both knowledge that the model has encountered and knowledge that remains novel to it.
Therefore, the existing methods face difficulties in handling the above situation.
On one hand, the present models rely entirely on Euclidean space throughout the training procedure. Consequently, when modeling samples that exhibit similar texts but dissimilar images, the prototypes representing different labels tend to become closer in proximity as shown in \figref{fig:introduction}.
On the other hand, the zero-shot learning based methods are limited to training solely with samples from seen categories, aiming to establish the semantic correlation between samples and their corresponding labels.
When generalized to a test set that encompasses both seen and unseen categories, the models tend to assign the sample as seen categories with high semantic similarities, while assigning low scores to unseen ones. Even with the application of a calibration factor to the seen categories, the models are unable to prevent the similarity distributions of all categories from becoming dispersed.
Due to the presence of a semantic similarity distribution disparity among unseen categories as illustrated in \figref{fig:introduction}, stemming from the absence of authentic training samples, this gap adversely affects the performance of generalized zero-shot learning approaches.

To overcome the above mentioned disadvantages, we propose the \textbf{h}yperbolic \textbf{m}ultimodal \textbf{g}enerative \textbf{r}epresentation \textbf{l}earning (HMGRL)~\footnote{\url{https://github.com/ZovanZhou/HMGRL}} framework to tackle the generalized zero-shot MIE (GZS-MIE) task.
Firstly, the hyperbolic variational information bottleneck (HVIB) is exploited to align the textual and visual modalities for bridging semantic gap between them.
Moreover, we design the hyperbolic multimodal conditional variational autoencoder (HMCVAE) and it can model the multimodal representations of samples and the correlation between them and label prototypes.
Based on the above two components, the framework can capture the multi-level hierarchical semantic correlations among samples and prototypes with the advantage of hyperbolic space.
During the training procedure, the prototypical knowledge of unseen categories is regarded as a condition to generate samples by the decoder of HMCVAE and the model is trained with real and synthetic samples.
To mitigate the numerical disparity in similarity scores inner samples of unseen categories, we introduce a semantic similarity distribution alignment loss, thereby enhancing the generalization capability of the GZS-MIE model.
The contributions of this manuscript can be summarized as follows:
\begin{itemize}
    \item We are the first one to investigate the generalized zero-shot multimodal information extraction (GZS-MIE) task and analyze the disadvantages of the existing ZS-IE and ZS-MIE models for this new task.
    \item We propose the hyperbolic multimodal generative representation learning (HMGRL) framework to tackle the GZS-MIE task.
    The framework does not only contain the HVIB and HMCVAE networks to capture the hierarchical semantic correlations of samples and prototypes but also employs synthetic samples to train the model with a semantic similarity distribution alignment loss, thereby gaining the better generalization ability.
    \item We carry out comprehensive experiments on two standard MIE benchmark datasets, and the results illustrate that our proposed framework outperforms the baseline methods.
\end{itemize}

\section{Related Work}

\subsection{Mulitmodal Information Extraction}
With the ever-increasing volume of multimodal data, researchers have come to realize the necessity of extracting the complex semantic information embedded within such data. As a result, they have expanded traditional information extraction (IE) tasks to include multimodal IE, aiming to build a multimodal knowledge graph~\cite{9961954}.
\citet{DBLP:conf/naacl/MoonNC18} first expanded the traditional realm of text-based named entity recognition to incorporate multimodal named entity recognition (MER). To tackle this task, the various fusion strategies including the modality attention mechanism~\cite{DBLP:conf/naacl/MoonNC18}, the unified multimodal transformer~\cite{yu-etal-2020-improving-multimodal} or the span-based variational autoencoder~\cite{zhou-etal-2022-span} were proposed to fuse textual and visual data, ultimately improving the performance on sequence label predictions.
\citet{DBLP:conf/mm/ZhengFFCL021} proposed the multimodal relation extraction (MRE) task, utilizing visual information to enhance the semantic understanding of textual content and accurately identify relations among entities. Furthermore, to jointly model the above two tasks, \citet{DBLP:conf/aaai/Yuan0WL23} introduced the multimodal entity-relation extraction (MERE) task and exploited the word-pair relation tagging strategy.
To ensure that the extracted multimodal entities are consistent with the human knowledge framework, \citet{zhang-etal-2023-incorporating-object} introduced the multimodal named entity typing (MET) task and employed a cross-modal transformer to analyze visual objects.

The above multimodal information extraction tasks are grounded in supervised learning paradigm, with the goal of directly mapping multimodal data to predefined labels. Nevertheless, our research centers on generalized zero-shot multimodal information extraction (GZS-MIE), which could identify not only seen but also unseen categories without the need for extra training. Generalized zero-shot learning emphasizes the creation of adaptable representations for both samples and semantic labels, facilitating the classification of samples into categories that have not been trained before.

\subsection{Zero-Shot Information Extraction}
Information extraction encompasses a range of tasks aimed at distilling structural information from unstructured texts for constructing knowledge graphs.
To enable the extraction of novel knowledge like: the identification of unseen categories including entity types or relations, without requiring further training, the zero-shot learning was introduced into IE and MIE tasks. The vital challenge for them is to learn generalizable representations of entities and prototypical knowledge of categories.
For zero-shot named entity typing (ZS-ET), the prototypical knowledge of types was encoded by the label embedding method and \citet{DBLP:conf/coling/MaCG16} introduced it to strengthen the semantic connection between the representations of entity mentions and their corresponding types.
In terms of data, \citet{DBLP:conf/naacl/ObeidatFST19,DBLP:conf/emnlp/0019JL0FYX21} incorporated supplementary information, such as descriptions of entity mentions and types gathered from websites, into the ZS-ET model to enhance the representations of both entity mentions and their corresponding types.
To extract the fine-grained features of contexts relevant to the types, the attention mechanism~\cite{DBLP:conf/www/RenLZ20} or the knowledge memory capabilities~\cite{DBLP:conf/coling/ZhangXLY20} were introduced into the ZS-ET models.
To address the challenge of zero-shot relation extraction (ZS-RE), entities and relations were input into a pre-trained language model, and \citet{chen-li-2021-zs} mapped them into an embedding space by simultaneously optimizing the semantic similarities among them.
\citet{DBLP:conf/acl/ZhaoZZZGWWPS23} and \citet{DBLP:conf/coling/GongE24a}
respectively proposed the semantic matching method and the prompt-based one for improving the performance on the ZS-RE task.
Except for the above text-based IE tasks, \citet{DBLP:conf/www/ZhouZZSY25} investigated the zero-shot multimodal information extraction (ZS-MIE) task and presented the fine-grained representation learning framework including the variational mixture of experts network and the multimodal graph-based virtual adversarial learning.

To sum up, existing ZS-IE and ZS-MIE methods are based on zero-shot learning. These approaches are limited to identifying samples exclusively from unseen categories and struggle to effectively recognize both seen and unseen categories simultaneously.
In contrast to these existing studies, our research centers on the GZS-MIE task, which seeks to extract base and novel structural knowledge embedded in multimodal data originating from social media platforms.

\section{Lorentz Linear Transformation}
We introduce the hyperbolic space to model the hierarchical semantic correlations of the samples and prototype representations.
There are two effective hyperbolic geometries including the Poincare ball~\cite{DBLP:conf/nips/GaneaBH18} and the Lorentz model~\cite{DBLP:conf/icassp/LiangWWBY25} which are utilized to describe hyperbolic space.
Considering the advantages of the Lorentz model~\cite{DBLP:conf/coling/LiangWBG24}, we exploit it to define the linear transformation operation in the hyperbolic space.
The $n$-dimensional Lorentz model $\mathbb{L}_{c}^n$ is the Riemannian manifold that satisfying: $\mathbb{L}_{c}^n = \{\textbf{x} \in \mathbb{R}^{n+1} | <\textbf{x},\textbf{x}>_{\mathbb{L}}=\frac{1}{c},x_0>0\}$,
where $c$ is the negative curvature, and $<\cdot,\cdot>_{\mathbb{L}}$ is the Lorentz scalar product.
Each point in the Riemannian manifold $\mathbb{L}_{c}^n$ has the form of $\textbf{x} = \left[x_0, \textbf{x}_s\right]$, where $x_0 \in \mathbb{R}$ and $\mathbf{x}_s \in \mathbb{R}^{n}$. The Lorentz scalar product of the point $\textbf{x}$ and \textbf{y} is defined as follows:
\begin{equation*}
    <\textbf{x},\textbf{y}>_{\mathbb{L}} = - x_0y_0 + \sum_{i=1}^n x_iy_i.
\end{equation*}
Therefore, the orthogonal space of the Riemannian manifold $\mathbb{L}_{c}^n$ at the point $\textbf{x}$ can be defined as: $\mathcal{T}_{\textbf{x}}\mathbb{L}_{c}^n = \{\textbf{y}\in\mathbb{R}^{n+1} | <\textbf{x},\textbf{y}>_{\mathbb{L}}=0\}$ where $\mathcal{T}_{\textbf{x}}\mathbb{L}_{c}^n$ is a Euclidean subspace.
And $\mathcal{T}_{0}\mathbb{L}_{c}^n$ represents the tangent space at the origin.

\begin{figure*}[t]
    \centering
    \includegraphics[width=\textwidth]{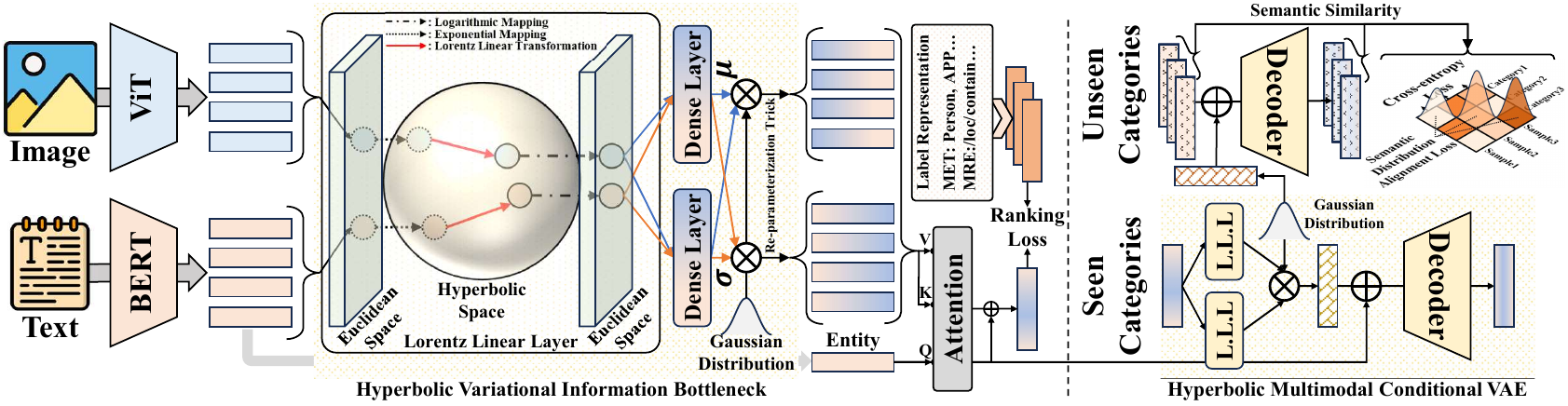}
    \setlength{\abovecaptionskip}{-0.2cm}
    \caption{
        The hyperbolic multimodal generative representation learning (HMGRL) framework for generalized zero-shot multimodal information extraction.
        The framework mainly contains the hyperbolic variational information bottleneck and hyperbolic multimodal conditional variational autoencoder (VAE).
        ``L.L.L'' is the abbreviation of the Lorentz linear layer.
        $\oplus$ represents the concatenation operation and $\otimes$ is the re-parameterization trick.
    }
    \label{fig:model}
    \vspace{-6pt}
\end{figure*}

The exponential mapping is utilized to map the feature vectors from the Euclidean space to the hyperbolic space.
To map the point $\textbf{x} \in \mathcal{T}_{\textbf{x}}\mathbb{L}_{c}^n$ to $\mathbb{L}_{c}^n$ by moving along the geodesic, the exponential map can be represented as:
\begin{equation*}
    \text{exp}_{\textbf{x}}^c (\textbf{z}) = \cosh (\alpha) \textbf{x} + \sinh (\alpha) \frac{\textbf{z}}{\alpha}
\end{equation*}
where $\alpha = \sqrt{-c} \|\textbf{z}\|_{\mathbb{L}}$ and $\|\textbf{z}\|_{\mathbb{L}} = \sqrt{<\textbf{z},\textbf{z}>_{\mathbb{L}}}$.
As the reversed mapping, the logarithmic map is exploited to map the vectors back to the Euclidean space. Given the point $\textbf{y} \in \mathbb{L}_c^n$ in hyperbolic space, the logarithmic map can be defined as:
\begin{equation*}
    \log_{\textbf{x}}^c (\textbf{y}) = \frac{\cosh^{-1}(\beta)}{\sqrt{\beta^2-1}} (\textbf{y} - \beta \textbf{x})
\end{equation*}
where $\beta = c<\textbf{x},\textbf{y}>_{\mathbb{L}}$.
Based on the above mapping functions, the Lorentz model provides the linear transformation from $\mathbb{L}_c^n$ to $\mathbb{L}_c^m$ for the two points $\textbf{x} \in \mathbb{L}_c^n$ and $\textbf{v} \in \mathcal{T}_{0}\mathbb{L}_{c}^n$.
Given a linear map $\mathcal{M}: \mathbb{R}^n \to \mathbb{R}^m$ with the weight matrix, the Lorentz linear transformation is defined as:
\begin{equation*}
    \mathcal{M}_c (\textbf{x}) = \exp_0^c (\hat{\mathcal{M}}(\log_0^c(\textbf{x}))), \hat{\mathcal{M}} = \left[0, \mathcal{M}(\textbf{v}_s)\right]
\end{equation*}
where $\textbf{v}_s = \left[v_1,v_2,\dots,v_n\right]$, $\exp_0^c: \mathcal{T}_{0}\mathbb{L}_{c}^n \to \mathbb{L}_c^m$ and $\log_0^c: \mathbb{L}_{c}^n \to \mathcal{T}_{0}\mathbb{L}_{c}^m$.
To model the hierarchical semantic correlations of samples and prototypes, we define the Lorentz linear layer to map the learned features into the hyperbolic space and then use the Lorentz linear transformation in the hyperbolic space.
Eventually, we utilize logarithm mapping to map the transformed features back to Euclidean space for tackling the MIE tasks.
The process of the \textbf{L}orentz \textbf{l}inear \textbf{l}ayer (LLL) is defined as follows:
\begin{equation*}
    \text{LLL}(\cdot;\mathcal{M}) = \log_0^c (\mathcal{M}_c(\exp_0^c(\cdot))) \in \mathbb{R}^m.
\end{equation*}

\section{Methodology}
Before introducing the details of the proposed model, we formalize the problem of generalized zero-shot multimodal information extraction (GZS-MIE).
We formalize the seen category set as $\mathcal{Y}_{s} = \{y_1^s, y_2^s, \dots, y_{|\mathcal{Y}_s|}^s\}$ with $|\mathcal{Y}_s|$ seen types and the unseen category set as $\mathcal{Y}_{u} = \{y_1^u, y_2^u, \dots, y_{|\mathcal{Y}_u|}^u\}$ with $|\mathcal{Y}_u|$ unseen types where $\mathcal{Y}_{s} \bigcap \mathcal{Y}_{u} = \emptyset$.
To train the model, we acquire the training dataset $\mathcal{D}_{train} = \{(T_i,V_i,E_i,Y_i)|Y_i \in \mathcal{Y}_{s},i=1,2,\dots,|\mathcal{D}_{train}|\}$ where $T$ denotes a natural language sentence, $V$ is the image incorporated with the sentence, $E$ represents the entity in the sentence, and $Y$ is the label for the MIE tasks.
Unlike the zero-shot learning, the test dataset of the generalized zero-shot learning is defined as $\mathcal{D}_{test} = \{(T_i,V_i,E_i,Y_i)|Y_i \in \mathcal{Y}_{s} \bigcup \mathcal{Y}_u,i=1,2,\dots,|\mathcal{D}_{test}|\}$.

In this section, we introduce the hyperbolic multimodal generative representation learning (HMGRL) framework for the GZS-MIE as shown in \figref{fig:model}.
In order to build up the framework, the details of it can be summarized as the following parts:
(1) To begin with, we utilize pre-trained language and vision models to extract multimodal input representations from the samples. And we exploit the hyperbolic variational information bottleneck as a unified backbone, aiming to align these multimodal input representations.
(2) Next, in order to capture the local semantic features inherent in the multimodal representations, we employ the attention mechanism and incorporate ranking loss to complete the GZS-MIE tasks.
(3) Moreover, we introduce a hyperbolic multimodal conditional variational autoencoder, designed to model the multimodal representations of samples and leverage the semantic relationships between samples and prototypes for generating synthetic samples belonging to unseen categories.
(4) Finally, to classify samples into seen and unseen categories, we compute the distances between the label semantic features and the multimodal features of the samples.

\subsection{Multimodal Representation Alignment}
Considering a quadruple $(T,V,E,Y)$ representing a multimodal sample that encompasses a text-image pair, we map the multimodal data into dense representations suitable for processing by deep neural networks.
For the textual modality, we represent the original sentence, which contains $|S|$ words, as $S=\{w_1, w_2, \dots, w_{|S|}\}$. To convert these discrete words into distributed representations, we utilize pre-trained language models, such as BERT, as the text encoder.
The extraction procedure for sentence features could be represented as $\textbf{T} = \text{BERT}(\tilde{S}) = \{\textbf{t}_{\texttt{[CLS]}}, \textbf{t}_1, \dots, \textbf{t}_{\texttt{[SEP]}}\} \in \mathbb{R}^{|T| \times d}$ where $\tilde{S}$ is the extended sentence with the special tokens and $|T|$ is the token number of it.
Regarding the visual modality, we employ the pre-trained vision transformer (ViT)~\cite{DBLP:conf/iclr/DosovitskiyB0WZ21} model to extract the features of an image that can be denoted as $\textbf{V} = \{\textbf{v}_1, \textbf{v}_2, \dots, \textbf{v}_{|V|}\} \in \mathbb{R}^{|V| \times d}$, where $|V|$ denotes the number of feature vectors obtained from it.

To bridge the semantic gap between the textual and visual modalities~\cite{DBLP:conf/wsdm/XuHSW22,DBLP:conf/nips/BaoW0LMASPW22}, we need to maximize the mutual information $\text{MI}(\textbf{T},\textbf{V})$ while reducing the noise information of them. Therefore, we propose to exploit the hyperbolic variational information bottleneck (HVIB) as the unified backbone to extract the effective features of them.
Given the textual representation $\textbf{T}$ and the visual one $\textbf{V}$, HVIB is trained to learn the latent variable $\textbf{Z}$ to preserve the sufficient features from two modalities for aligning them.
Considering the symmetry of multimodal representation alignment, the information bottlenecks~\cite{DBLP:conf/itw/TishbyZ15} for the two modalities are defined as:
\begin{equation}
\begin{aligned}
    \mathcal{L}_{IB}^{T\to V} & = \beta \cdot \text{MI}(\textbf{T},\textbf{Z}^{T \to V}) - \text{MI}(\textbf{Z}^{T \to V},\textbf{V}), \\
    \mathcal{L}_{IB}^{V\to T} & = \beta \cdot \text{MI}(\textbf{V},\textbf{Z}^{V \to T}) - \text{MI}(\textbf{Z}^{V \to T},\textbf{T}).
\end{aligned}
\end{equation}
We minimize the items $\text{MI}(\textbf{T},\textbf{Z}^{T \to V})$ and $\text{MI}(\textbf{V},\textbf{Z}^{V \to T})$ to reduce the noise information from the two modalities.
And to ensure the semantic alignment of them inner a sample, we maximizes the items $\text{MI}(\textbf{Z}^{T \to V},\textbf{V})$ and $\text{MI}(\textbf{Z}^{V \to T},\textbf{T})$ by predicting their aligned modality features.
Then, we exploit the variational manner to encode the latent representations and the gaussian variables $\textbf{Z}^{T \to V}$ and $\textbf{Z}^{V \to T}$ are calculated with the re-parameterization trick~\cite{DBLP:journals/corr/KingmaW13} as follows:
\begin{equation}
    \textbf{Z}^{T \to V} = \bm{\mu}_{T \to V} + \bm{\sigma}_{T \to V} \odot \bm{\epsilon}, \quad \textbf{Z}^{V \to T} = \bm{\mu}_{V \to T} + \bm{\sigma}_{V \to T} \odot \bm{\epsilon}
\end{equation}
where $\bm{\epsilon} \sim \mathcal{N}(0,\textbf{I})$ and $\odot$ is the element-wise multiplication operation.
To model the hierarchical semantic correlation between the two modalities, we define the calculation procedure for the mean variable $\bm{\mu}$ and standard deviation one $\bm{\sigma}$ as:
\begin{equation}
\begin{aligned}
    \bm{\mu}_{T \to V} = \psi(\text{LLL}(\textbf{T};\mathcal{M}_{\mu});\bm{\theta}_{\mu}), \quad \bm{\sigma}_{T \to V} = \psi(\text{LLL}(\textbf{T};\mathcal{M}_{\sigma});\bm{\theta}_{\sigma})\\
    \bm{\mu}_{V \to T} = \psi(\text{LLL}(\textbf{V};\mathcal{M}_{\mu});\bm{\theta}_{\mu}), \quad \bm{\sigma}_{V \to T} = \psi(\text{LLL}(\textbf{V};\mathcal{M}_{\sigma});\bm{\theta}_{\sigma})
\end{aligned}
\end{equation}
where $\psi(\cdot;\theta)$ represents the feed-forward neural network with the parameter matrix $\theta$ and $\{\mathcal{M}_\mu,\mathcal{M}_\sigma\}$ are the weight matrices in the Lorentz linear layers.
To optimize the model by the IB principle, we can use the variational posterior distributions $q(\textbf{Z}^{T \to V}|\textbf{T})$ and $q(\textbf{Z}^{V \to T}|\textbf{V})$ to approximate the posterior distributions $p(\textbf{Z}^{T \to V}|\textbf{T})$ and $p(\textbf{Z}^{V \to T}|\textbf{V})$~\cite{DBLP:journals/taslp/CuiCCSLLS24}.
By assuming the prior distributions $p(\textbf{Z}^{T \to V})$ and $p(\textbf{Z}^{V \to T})$ as the standard normal distribution, the regularization loss for HVIB is defined as follows:
\begin{equation} \label{eqn:reg_loss}
\begin{aligned}
    \mathcal{L}_{reg} = &\frac{1}{2} (\text{KL}(\mathcal{N}(\bm{\mu}_{T \to V}, \bm{\sigma}_{T \to V}^2)||\mathcal{N}(0,\textbf{I})) \\
     &+ \text{KL}(\mathcal{N}(\bm{\mu}_{V \to T}, \bm{\sigma}_{V \to T}^2)||\mathcal{N}(0,\textbf{I}))).
\end{aligned}
\end{equation}
Moreover, to maximize the items $\text{MI}(\textbf{Z}^{T \to V},\textbf{V})$ and $\text{MI}(\textbf{Z}^{V \to T},\textbf{T})$, we can directly maximize $\text{MI}(\textbf{Z}^{T \to V},\textbf{Z}^{V \to T})$ by the contrastive learning~\cite{DBLP:conf/aaai/ZareapoorSL25} for aligning the two modalities.
Given a batch of $N$ samples, we apply the average pooling on their latent representations and acquire the global ones $\{(\bar{\textbf{Z}}^{T \to V}_i, \bar{\textbf{Z}}^{V \to T}_i)|i=1,2,\dots,N\}$ where $\bar{\textbf{Z}}^{T \to V}_i=\frac{1}{|T|}\sum_{j=1}^{|T|}\textbf{z}^{T \to V}_{i,j}$ and $\bar{\textbf{Z}}^{V \to T}_i=\frac{1}{|V|}\sum_{j=1}^{|V|}\textbf{z}^{V \to T}_{i,j}$.
Contrastive learning aims to distinguish matched pairs from the $N \times N$ possible combinations of image and text latent representations, making sure that the representations of paired inputs are closer to each other in the representation space than those of unpaired inputs.
Therefore, the contrastive learning loss is defined as follows:
\begin{equation} \label{eqn:cl_loss}
\begin{aligned}
    \mathcal{L}_{cl} = \sum_{i=1}^{N}- & \frac{1}{2} (\log \frac{\exp(\bar{\textbf{Z}}_i^{T \to V} \cdot \bar{\textbf{Z}}_i^{V \to T})}{ \sum_{j=1}^N \exp(\bar{\textbf{Z}}_i^{T \to V} \cdot \bar{\textbf{Z}}_j^{V \to T})} \\
    & + \log \frac{\exp(\bar{\textbf{Z}}_i^{V \to T} \cdot \bar{\textbf{Z}}_i^{T \to V})}{ \sum_{j=1}^N \exp(\bar{\textbf{Z}}_i^{V \to T} \cdot \bar{\textbf{Z}}_j^{T \to V})}).
\end{aligned}
\end{equation}

\subsection{Generalized Zero-Shot MIE}
To complete the MIE tasks, we mark the begin and end of the entities according to the entity set $E$ and insert the reserved tokens \texttt{[E1]}, \texttt{[/E1]} (and \texttt{[E2]}, \texttt{[/E2]}) into the original sentence~\cite{zhou-etal-2024-mcil}.
Then, we extract the entity representation according to the specific tasks.
For the different tasks, we denote the entity representation as:
\begin{equation}
\begin{aligned}
\text{MET:} \quad & \textbf{e} = [\textbf{t}_{\texttt{[CLS]}} \oplus \textbf{t}_{\texttt{[E1]}}] \in \mathbb{R}^{2d} \\
\text{MRE:} \quad & \textbf{e} = [\textbf{t}_{\texttt{[CLS]}} \oplus \textbf{t}_{\texttt{[E1]}} \oplus \textbf{t}_{\texttt{[E2]}}] \in \mathbb{R}^{3d}
\end{aligned}
\end{equation}
where $\textbf{t}_{\texttt{[CLS]}}$, $\textbf{t}_{\texttt{[E1]}}$ and $\textbf{t}_{\texttt{[E2]}}$ represent the textual representations of the special tokens \texttt{[CLS]}, \texttt{[E1]} and \texttt{[E2]} respectively, and $\oplus$ is the vector concatenation operation.
To acquire the fusion multimodal representation, we concatenate the aligned latent representations as the fusion one $\textbf{U} = [\textbf{Z}^{T \to V}, \textbf{Z}^{V \to T}] \in \mathbb{R}^{(|T|+|V|) \times d}$.
For extracting the local features of the fusion multimodal representation relevant to the entity, we apply the cross-modal attention mechanism on the features of different modalities.
With the use of textual entity feature $\textbf{e} \in \mathbb{R}^{|\textbf{e}|}$ as the query and the fusion multimodal representation $\textbf{U} = \{\textbf{u}_i|i=1,2,\dots,|T|+|V|\}$ as the key and value, the attention score is calculated as $\beta_i = \frac{\phi([\textbf{u}_i\oplus\textbf{e}])}{\sum_{j=1}^{|T|+|V|} \phi( [\textbf{u}_j\oplus\textbf{e}])}$
where $\phi(\textbf{x}) = \exp(\textbf{W}_A \cdot \textbf{x} + \textbf{b}_A)$, and $\textbf{W}_A \in \mathbb{R}^{(|\textbf{e}|+d)}$ and $\textbf{b}_A \in \mathbb{R}$ are the trainable parameters in the cross-modal attention layer.
Therefore, the entity-aware multimodal fusion representation is defined as $\bar{\textbf{U}} = \sum_{i=1}^{|T|+|V|} \beta_i \textbf{u}_i$.
Then, we concatenate the above feature and textual entity one as the whole $\tilde{\textbf{e}} = [\bar{\textbf{U}} \oplus \textbf{e}] \in \mathbb{R}^{(|\textbf{e}|+d)}$.

To identify the category under the generalized zero-shot learning, we utilize the label embedding method to encode the names of labels as the prototype representation.
Given the seen category set $\mathcal{Y}_{s} = \{y_1^s, y_2^s, \dots, y_{|\mathcal{Y}_s|}^s\}$, we regard each label name as a short sentence and use BERT to encode it for acquiring its textual representations $\textbf{p}^s_i$.
Then, we apply the average pooling on it and the semantic representation of the category is denoted as $\bar{\textbf{p}}^s_i$.
Therefore, the prototype representations of the seen category set can be represented as $\textbf{P}^s = [\bar{\textbf{p}}^s_1, \bar{\textbf{p}}^s_2, \dots, \bar{\textbf{p}}^s_{|\mathcal{Y}_s|}] \in \mathbb{R}^{|\mathcal{Y}_s| \times d}$.
To assess the correlation between the sample and categories, we employ the semantic similarity as $\textbf{O}^s = \psi(\tilde{\textbf{e}};\bm{\theta}_{e}) \psi(\textbf{P}^s;\bm{\theta}_{p})^T \in \mathbb{R}^{|\mathcal{Y}_s|}$.
To ensure the similarity of the true label keeps higher than those of other labels, we optimize the model by leveraging the ranking loss as:
\begin{equation} \label{eqn:ranking_loss}
    \mathcal{L}_{rank} = \sum_{i=1}^{|\mathcal{Y}_s|} \max (1 - o^s_{+} + o_i^s, 0)
\end{equation}
where $o_+^s$ is the similarity between the sample and the true type.

\begin{table*}[ht]
    \caption{
        Performance comparison on the MET and MRE datasets under the generalized zero-shot learning settings.
        The \textbf{bolded numbers} indicate the best results and the \underline{underlined numbers} represent the second-best results.
    }
    \label{tab:main-results}
    \resizebox{\textwidth}{!}{
    \begin{tabular}{lcccccc|lcccccc}
    \toprule
    \multicolumn{7}{c|}{Multimodal Named Entity Typing}                                                     & \multicolumn{7}{c}{Multimodal Relation Extraction}                                                     \\ \midrule
    \multirow{2}{*}{Model} & \multicolumn{2}{c}{Seen} & \multicolumn{2}{c}{Unseen} & \multicolumn{2}{c|}{Overall} & \multirow{2}{*}{Model} & \multicolumn{2}{c}{Seen} & \multicolumn{2}{c}{Unseen} & \multicolumn{2}{c}{Overall} \\
                           & Accuracy         & F1         & Accuracy          & F1          & Accuracy        & F1        &                        & Accuracy         & F1         & Accuracy          & F1          & Accuracy       & F1        \\ \midrule
    Proto                  & \underline{67.9$\pm$27.2}   & \underline{73.3$\pm$23.5}  & 15.3$\pm$10.0    & 14.2$\pm$6.4    & 21.9$\pm$9.0   & 22.5$\pm$8.9  & Proto                  & 43.6$\pm$3.3    & 51.0$\pm$5.3   & \underline{31.6$\pm$18.5}    & \underline{29.6$\pm$15.5}   & 34.6$\pm$15.2 & 36.2$\pm$14.9 \\
    DBZFET                 & \textbf{76.7$\pm$7.1}    & \textbf{82.4$\pm$7.2}   & 19.8$\pm$14.2    & 20.8$\pm$14.1   & 29.4$\pm$15.7  & 31.2$\pm$16.5 & ZS-BERT                & 53.1$\pm$15.3   & 59.7$\pm$15.3  & 17.3$\pm$2.8     & 20.7$\pm$3.0    & 25.7$\pm$3.4  & 30.6$\pm$4.9  \\
    NZFET                  & 63.3$\pm$32.8   & 66.2$\pm$33.1  & 7.4$\pm$5.6      & 7.5$\pm$5.2     & 11.6$\pm$7.6   & 12.9$\pm$8.8  & RE-match            & 49.9$\pm$6.8    & 56.0$\pm$4.8   & 26.9$\pm$16.3    & 28.7$\pm$14.1   & 31.9$\pm$14.7 & 36.3$\pm$13.4 \\
    MZET                   & 19.8$\pm$18.9   & 20.4$\pm$19.4  & \underline{29.1$\pm$20.0}    & \underline{31.2$\pm$20.6}   & 14.5$\pm$11.4  & 16.1$\pm$13.0 & ZS-SKA                 & \underline{53.7$\pm$8.8}    & \underline{60.3$\pm$7.2}   & 24.6$\pm$12.9    & 23.6$\pm$14.6   & 32.6$\pm$14.3 & 32.3$\pm$16.9 \\
    MMProto                & 62.4$\pm$30.9   & 68.9$\pm$29.2  & 18.2$\pm$11.3    & 21.6$\pm$11.9   & 24.0$\pm$12.9  & 30.3$\pm$15.3 & MMProto                & \textbf{54.8$\pm$0.8}    & \textbf{62.5$\pm$1.1}   & 29.1$\pm$14.8    & 28.8$\pm$10.8   & \underline{36.5$\pm$13.0} & \underline{38.6$\pm$10.8} \\
    MOVCNet                & 65.5$\pm$32.8   & 68.6$\pm$34.4  & 14.0$\pm$3.2     & 15.5$\pm$8.5    & 21.5$\pm$2.9   & 25.2$\pm$13.6 & MOVCNet                & 35.1$\pm$8.6    & 42.6$\pm$3.4   & 31.2$\pm$9.1     & 27.4$\pm$6.1    & 31.5$\pm$2.7  & 32.9$\pm$3.9  \\
    MG-VMoE                & 55.7$\pm$25.8   & 61.6$\pm$28.1  & 25.3$\pm$21.4    & 29.4$\pm$25.1   & \underline{30.2$\pm$20.8}  & \underline{36.0$\pm$24.4} & MG-VMoE                & 32.4$\pm$6.8    & 37.5$\pm$11.1  & 26.4$\pm$18.7    & 26.5$\pm$16.2   & 26.1$\pm$14.6 & 30.1$\pm$14.7 \\
    Ours                   & 62.3$\pm$19.6   & 68.8$\pm$24.7  & \textbf{33.4$\pm$3.9}     & \textbf{38.8$\pm$2.6}    & \textbf{42.8$\pm$7.3}  & \textbf{48.2$\pm$7.3}  & Ours                   & 49.5$\pm$6.7    & 55.4$\pm$5.0   & \textbf{37.9$\pm$12.6}    & \textbf{35.5$\pm$14.7}   & \textbf{42.6$\pm$10.8} & \textbf{42.4$\pm$12.7} \\ \hline
    LLaVA                  & 54.6$\pm$12.7   & 61.2$\pm$17.0  & 37.8$\pm$31.1    & 41.6$\pm$23.8   & 57.7$\pm$15.7  & 49.5$\pm$16.2 & LLaVA                  & 20.6$\pm$3.6    & 18.4$\pm$8.2   & 12.2$\pm$3.7     & 15.2$\pm$1.7    & 14.9$\pm$2.1  & 15.7$\pm$3.1  \\
    \bottomrule
    \end{tabular}
    }
    % \vspace{-12pt}
\end{table*}

\subsection{Hyperbolic Multimodal Conditional Variational Autoencoder}
Considering that there are no samples from the unseen category set, we need to bridge the domain gap between the seen and unseen categories.
To take advantage of the generative model, we exploit it to generate the synthetic samples for adapting the model to the unseen category domain.
Therefore, we propose the hyperbolic multimodal conditional variational autoencoder (HMCVAE) to model the semantic correlation between the textual prototypes and samples' multimodal representations.
Compared with the traditional VAE model, the HMCVAE model consists of the encoder that is composed of the Lorentz linear layers for modeling the hierarchical semantic correlation between the samples and prototypes.
Given the multimodal representation $\tilde{\textbf{e}}^s$ of a sample from the seen category set, we can encode the latent representations for the HMCVAE model by the re-parameterization trick as follows:
\begin{equation}
\begin{aligned}
    \textbf{Z}^{VAE} &= \bm{\mu}_{VAE} + \bm{\sigma}_{VAE} \odot \bm{\epsilon};\\
    \bm{\mu}_{VAE} = \text{LLL}(\tilde{\textbf{e}}^s;&\mathcal{M}_{\mu}^{\prime}); \quad \bm{\sigma}_{VAE} = \text{LLL}(\tilde{\textbf{e}}^s;\mathcal{M}_{\sigma}^{\prime})
\end{aligned}
\end{equation}
where $\bm{\epsilon} \in \mathbb{R}^h$ and $h$ is the dimension number of the latent representation.
Considering to model the multimodal representation of the sample from the seen category set, we exploit the entity representation as the conditional prototypical knowledge for the different MIE tasks as follows:
\begin{equation}
\begin{aligned}
\text{MET:} \quad & \textbf{p}^e = \textbf{t}_{\texttt{[E1]}} \in \mathbb{R}^d \\
\text{MRE:} \quad & \textbf{p}^e = \frac{1}{2} (\textbf{t}_{\texttt{[E1]}} + \textbf{t}_{\texttt{[E2]}}) \in \mathbb{R}^d.
\end{aligned}
\end{equation}
Then, we feed the latent representation and the conditional prototypical one into the decoder to reconstruct the multimodal one of the input sample as $\hat{\textbf{e}}^s = \psi([\textbf{p}^e \oplus \textbf{Z}^{VAE}];\bm{\theta}_{d}) \in \mathbb{R}^{(|\textbf{e}|+d)}$ where $\bm{\theta}_d$ represents the set of weight parameters in the decoder.
The variational evidence lower bound (ELBO)~\cite{DBLP:journals/corr/KingmaW13} of HMCVAE could be defined as follows:
\begin{equation}
    \text{ELBO}(\tilde{\textbf{e}}^s) = \text{KL}(q(\textbf{Z}^{VAE}|\tilde{\textbf{e}}^s)||p(\textbf{Z}^{VAE})) + \log p(\tilde{\textbf{e}}^s|\textbf{Z}^{VAE})
\end{equation}
where the variational posterior distribution $q(\textbf{Z}^{VAE}|\tilde{\textbf{e}}^s)$ is the Gaussian distribution with the parameters $\{\bm{\mu}_{VAE},\bm{\sigma}_{VAE}\}$ and the prior one $p(\textbf{Z}^{VAE})$ is the standard normal distribution.
Therefore, the training loss for the HMCVAE is defined as the following function:
\begin{equation}
    \mathcal{L}_{vae} = \text{KL}(\mathcal{N}(\bm{\mu}_{VAE},\bm{\sigma}_{VAE}^2)||\mathcal{N}(0,\textbf{I})) + ||\tilde{\textbf{e}}^s - \hat{\textbf{e}}^s||^2.
\end{equation}

With the generative ability of HMCVAE, we can make use of the decoder of it to generate the synthetic samples.
Given the unseen category set $\mathcal{Y}_u$, we also utilize the label embedding method to encode the prototype representations of them and they can be denoted as $\textbf{P}^u = [\bar{\textbf{p}}^u_1, \bar{\textbf{p}}^u_2, \dots, \bar{\textbf{p}}^u_{|\mathcal{Y}_u|}] \in \mathbb{R}^{|\mathcal{Y}_u| \times d}$.
Through sampling the noise vector $\bm{\epsilon}$ from the standard normal distribution, we concatenate it and the prototype representation of the $i$-th unseen category to generate the synthetic sample as $\hat{\textbf{e}}^u_i = \psi([\bar{\textbf{p}}^u_i \oplus \bm{\epsilon}];\bm{\theta}_d) \in \mathbb{R}^{(|\textbf{e}|+d)}$.
Therefore, the synthetic multimodal representations of the unseen category set could be defined as $\hat{\textbf{E}}^u = [\hat{\textbf{e}}^u_1,\hat{\textbf{e}}^u_2,\dots,\hat{\textbf{e}}^u_{|\mathcal{Y}_u|}] \in \mathbb{R}^{|\mathcal{Y}_u| \times (|\textbf{e}|+d)}$.
Then, we can calculate the semantic similarities between the synthetic samples and unseen categories as $\textbf{O}^u = \psi(\hat{\textbf{E}}^u;\bm{\theta}_e)\psi(\textbf{P}^u;\bm{\theta}_p)^T \in \mathbb{R}^{|\mathcal{Y}_u| \times |\mathcal{Y}_u|}$.
To improve the performance of the model on identifying the unseen categories, we optimize the model with the cross-entropy loss as follows:
\begin{equation} \label{eqn:ce_loss}
    \mathcal{L}_{ce} = - \sum_{i=1}^{|\mathcal{Y}_u|} y^u_i \log(\text{softmax} (\textbf{o}_i^u))
\end{equation}
where $\textbf{o}_i^u \in \mathbb{R}^{|\mathcal{Y}_u|}$ is semantic similarity vector of the $i$-th synthetic sample.
Considering that the cross-entropy loss cannot guarantee the close similarity score distributions cross different categories, the category with the low mean similarity score will not be identified efficiently.
To reduce the gap between the scores of the unseen categories, we propose the semantic distribution alignment loss as the following function:
\begin{equation} \label{eqn:align_loss}
    \mathcal{L}_{align} = \text{KL}(\text{softmax}(\text{Diag}(\textbf{O}^u))||\text{softmax}(\text{Diag}(\textbf{A})))
\end{equation}
where $\textbf{A} \in \mathbb{R}^{|\mathcal{Y}_u| \times |\mathcal{Y}_u|}$ is the identity matrix and $\text{Diag}(\cdot)$ represents to extract elements on the diagonal of the matrix as a vector. 

\subsection{Training and Inference Procedure}
To train the HMGRL framework with all loss functions in an end-to-end way, we introduce the hyper-parameters to sum the different losses.
Therefore, the overall loss function is defined as follows:
\begin{equation} \label{eqn:overall_loss}
    \mathcal{L}_{overall} = \mathcal{L}_{reg} + \mathcal{L}_{cl} + \mathcal{L}_{rank} + \mathcal{L}_{vae} + \eta \cdot \mathcal{L}_{ce} + \zeta \cdot \mathcal{L}_{align}
\end{equation}
where $\eta$ and $\zeta$ are the hyper-parameters to balance the losses related to the unseen category set.
Subsequently, we employ stochastic gradient descent (SGD)~\cite{DBLP:journals/corr/KingmaB14} algorithm to optimize the model weights based on the loss value calculated by \eqref{eqn:overall_loss}.

The GZS-MIE methods typically exhibit a bias towards the seen categories.  Because they can only employ samples from the seen category set for training and then are evaluated on both seen and unseen categories. This bias consequently results in the misclassification of data from unseen categories as belonging to the seen categories.
During the inference procedure, we introduce the calibration factor~\cite{DBLP:conf/cvpr/Liu00H00H21} to reduce the above bias.
Given a sample from the test set $\mathcal{D}_{test}$, we calculate the semantic similarity between it and the all categories as $\hat{\textbf{O}} =  \psi(\tilde{\textbf{e}};\bm{\theta}_{e}) \psi([\textbf{P}^s \oplus \textbf{P}^u];\bm{\theta}_{p})^T \in \mathbb{R}^{(|\mathcal{Y}_s| + |\mathcal{Y}_u|)}$.
With the calibration stacking method~\cite{DBLP:conf/eccv/ChaoCGS16}, we predict the category of the sample as the following function:
\begin{equation}
    \hat{y} = \underset{y_i \in \mathcal{Y}_s \bigcup \mathcal{Y}_u}{\arg \max} \hat{\textbf{o}}_i - \gamma \cdot \mathbb{I}(y_i \in \mathcal{Y}_s)
\end{equation}
where $\gamma$ is the calibration factor and $\mathbb{I}(\cdot) = 1$ if $y_i$ belongs to the seen category set and $0$ otherwise.

\section{Experiments}

\subsection{Datasets and Experiment Settings}
We investigate the generalized zero-shot multimodal information extraction, and undertake experiments on the two tasks: multimodal named entity typing (MET) and multimodal relation extraction (MRE).
In the MET task, we employ WikiDiverse~\cite{wang-etal-2022-wikidiverse} as the benchmark dataset, where each sample consists of a text-image pair from Wikinews, with entity mentions annotated into 13 fine-grained types. For the MRE task, we use the dataset introduced by \citet{DBLP:conf/mm/ZhengFFCL021}, which is constructed from Twitter posts. Annotators selected samples covering various topics, yielding a dataset with 23 relation types. Since both datasets contain non-semantic categories (e.g., ``Other'' or ``None''), we exclude them and retain only semantically meaningful categories~\cite{DBLP:conf/www/ZhouZZSY25}.

To emulate generalized zero-shot learning conditions~\cite{DBLP:conf/naacl/ZhangZLY25}, we randomly divide the original category set into three disjoint subsets. For the MET task, 4 categories are assigned to each of the training, validation, and test sets. For the MRE task, 8, 7, and 7 categories are allocated to the training, validation, and test sets, respectively. Subsequently, we randomly select 70\% of instances from each seen category. Of these, 90\% are placed in the training set, and the remaining 10\% are designated for the validation set. The test set includes the remaining 30\% of instances from seen categories and all instances from unseen categories.
The dimensionality $d$ of hidden representations and the learning rate are set to 768 and 1e-5 respectively.
We configure 20 epochs and a batch size of 14 for MET, and 25 epochs with a batch size of 8 for MRE.
The dimensionality $h$ of latent representations undergoes search across $\{256, 512, 768, 1024\}$, whereas hyper-parameters $\eta$ and $\zeta$ are searched in $\{1, 5, 10, 30, 50\}$. For $\gamma$ selection, we evaluate the model on the validation set following each training epoch.
We run each experiment three times with different random seeds and report the mean and standard deviation of performance metrics.
All experiments are expedited using NVIDIA GTX A6000 devices.
The compared methods and evaluation metrics are described in Appendix \ref{appendix:compared_methods}.

\begin{figure}[t]
    \centering
    \includegraphics[width=\columnwidth]{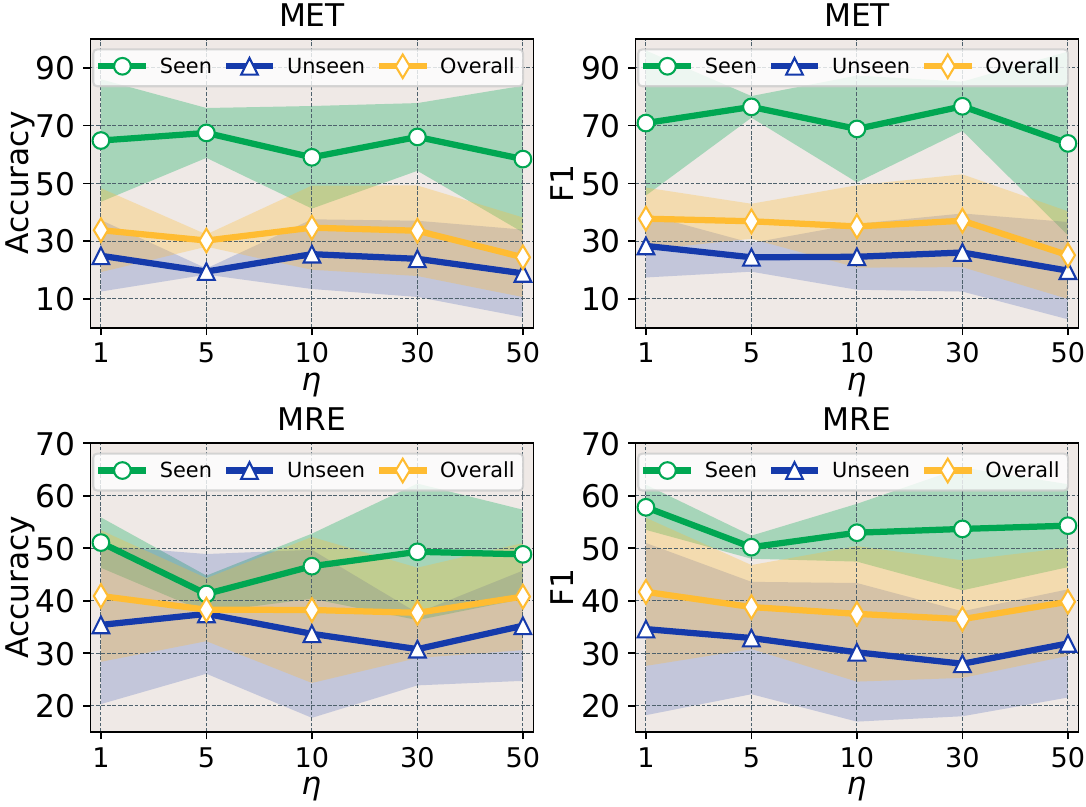}
    \setlength{\abovecaptionskip}{-0.2cm}
    \caption{
        A comparative analysis to the performance of HMGRL across different hyper-parameter $\eta$ values.
    }
    \label{fig:eta_results}
    \vspace{-6pt}
\end{figure}

\subsection{Experimental Results}
The HMGRL framework is evaluated alongside baseline models on the MET and MRE benchmark datasets, and the detailed experimental results are shown in \tbref{tab:main-results}.
Multimodal-based methods outperform text-based ones across most evaluation metrics on the two tasks, thereby confirming the efficacy of incorporating visual modalities.
Our model achieves the highest performance in terms of both accuracy and F1 scores on the ``Unseen'' and ``Overall'' items of the MET dataset, surpassing the second-best results by 12.6\% and 12.2\% for overall F1 and accuracy, respectively.
\begin{figure}[t]
    \centering
    \includegraphics[width=\columnwidth]{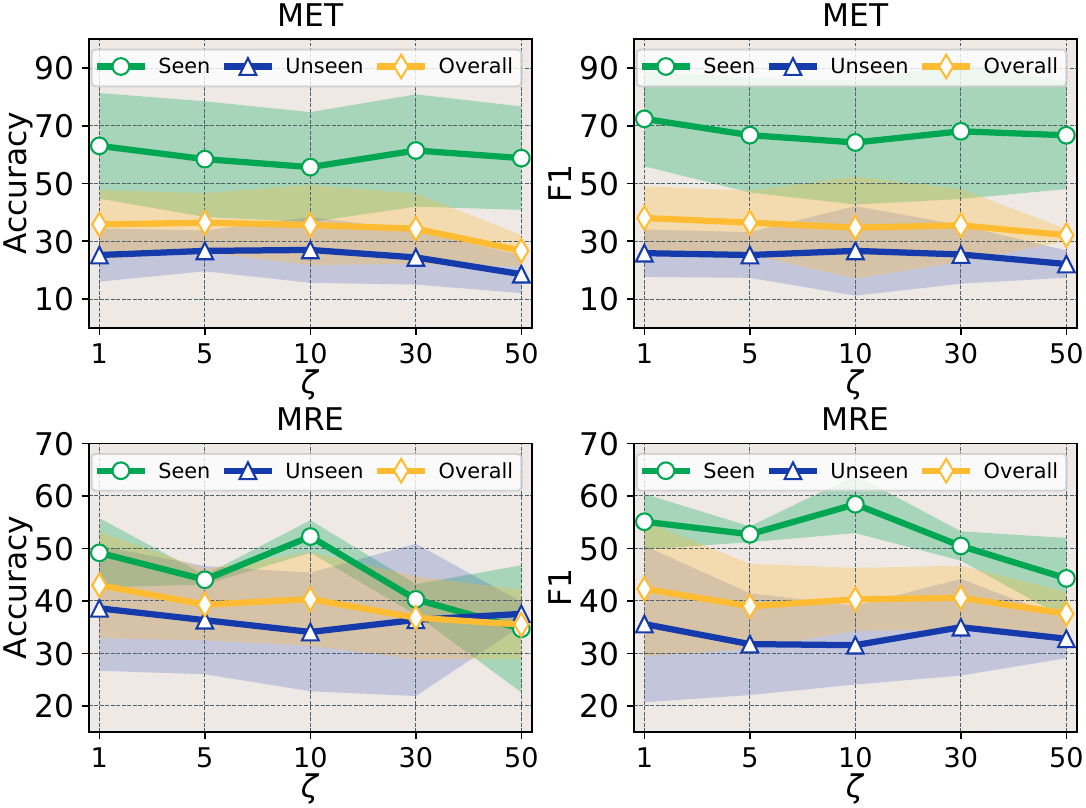}
    \setlength{\abovecaptionskip}{-0.2cm}
    \caption{
        A comparative analysis to the performance of HMGRL across different hyper-parameter $\zeta$ values.
    }
    \label{fig:zeta_results}
    \vspace{-6pt}
\end{figure}
\begin{table}[t]
    \setlength{\abovecaptionskip}{0.3cm}
    \caption{
        The ablation study results of HMGRL on the MET and MRE benchmark datasets.
    }
    \label{tab:ablation-study}
    \begin{tabular}{l|cc|cc}
        \toprule
        \multirow{2}{*}{Model}   & \multicolumn{2}{c|}{MET} & \multicolumn{2}{c}{MRE} \\
                                 & Accuracy       & F1     & Accuracy       & F1     \\ \midrule
        HMGRL                  & 37.7     & 40.3         & 42.6     & 42.4         \\
        ~~w/o HVIB               & 23.5     & 32.5         & 35.7     & 37.2         \\
        ~~w/o $\mathcal{L}_{cl}$             & 34.5     & 33.9         & 33.4     & 37.8         \\ 
        ~~w/o $\mathcal{L}_{vae}$ & 22.1 & 27.5&27.1&30.1 \\
        ~~w/o $\mathcal{L}_{ce}$ &26.0&30.4&35.3&36.0 \\
        ~~w/o $\mathcal{L}_{align}$ &36.2&35.6&39.8&39.8 \\ \bottomrule
    \end{tabular}
    \vspace{-6pt}
\end{table}
The aforementioned findings also underscore the efficacy of the proposed framework in identifying novel entity types although it entails a minor reduction in performance for the seen categories.
On the MRE dataset, our model surpasses the baseline methods in both accuracy and F1 scores of the ``Unseen'' and ``Overall'' items, attaining an accuracy improvement of 6.1\% and an F1 score enhancement of 3.8\% over the second-best performance.
When it comes to employing a large language model for zero-shot inference without any training, LLaVA demonstrates superior performance compared to our model on the MET dataset; however, it falls short on the MRE dataset.
Because the backbone model LLaMA~\cite{DBLP:journals/corr/abs-2302-13971} of LLaVA was trained with the Wikipedia pages and the entities in the MET dataset were stored on them.
To sum up, our model surpasses both text-based and multimodal baselines, thanks to the utilization of multimodal representation alignment facilitated by the HVIB network, along with the development of HMCVAE. This approach effectively models the multimodal representations of samples and generates synthetic data, ultimately enhancing the performance.

\begin{figure*}[t]
    \centering
    \subfloat[The ablation results on the Lorentz linear layer in HVIB and HMCVAE.]{\label{fig:lll_ablation_study}\includegraphics[width=0.32\textwidth]{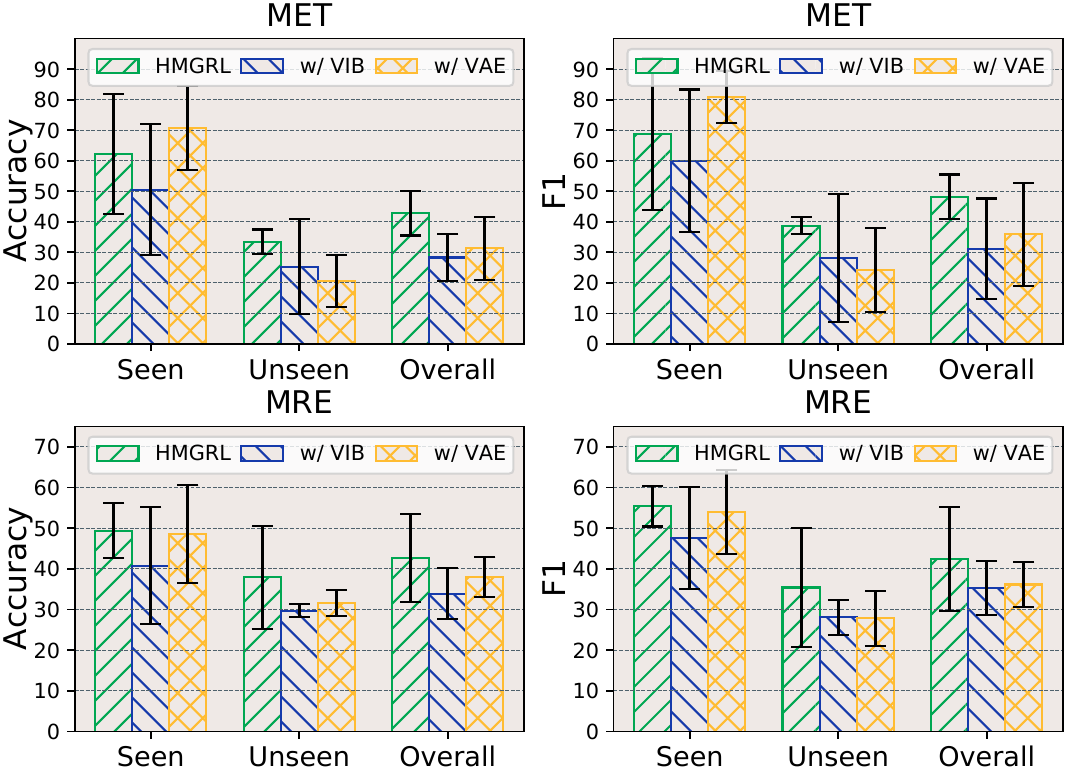}} \hspace{3pt}
    \subfloat[Performance comparison of HMGRL across different dimensionalities ($h$) of HVIB's latent representation.]{\label{fig:lll_vib_dim}\includegraphics[width=0.32\textwidth]{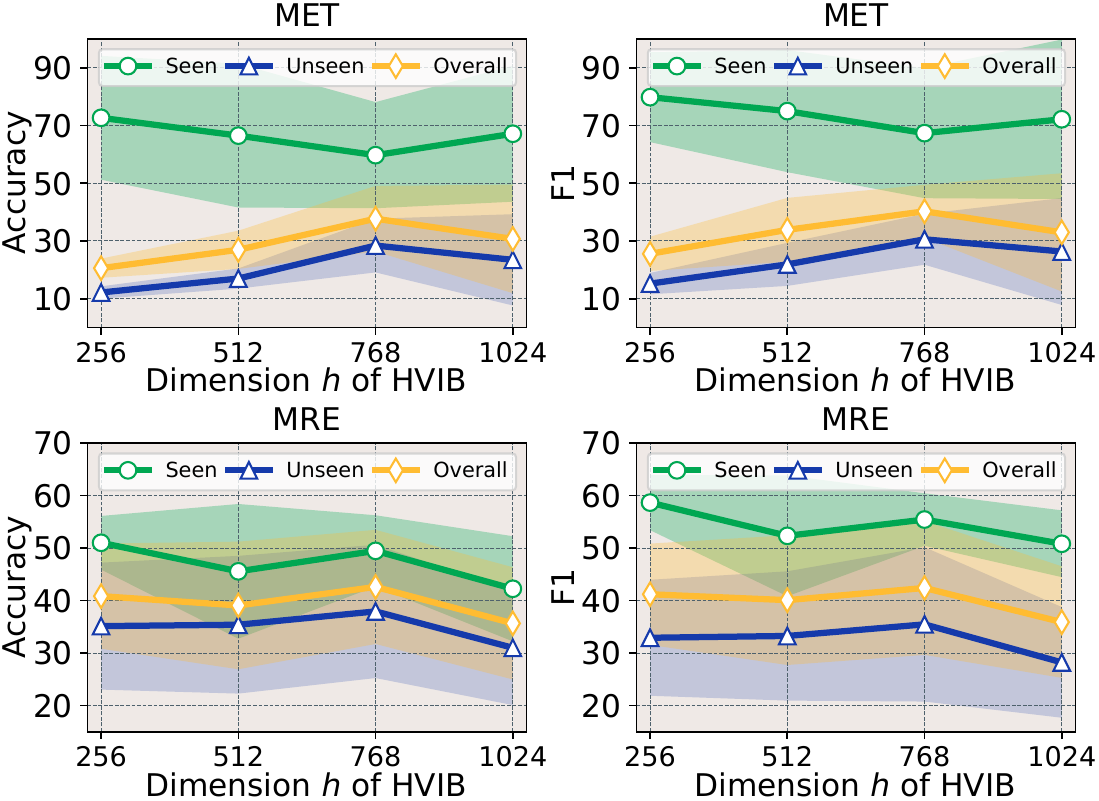}} \hspace{3pt}
    \subfloat[Performance comparison of HMGRL across different dimensionalities ($h$) of HMCVAE's latent representation.]{\label{fig:lll_vae_dim}\includegraphics[width=0.32\textwidth]{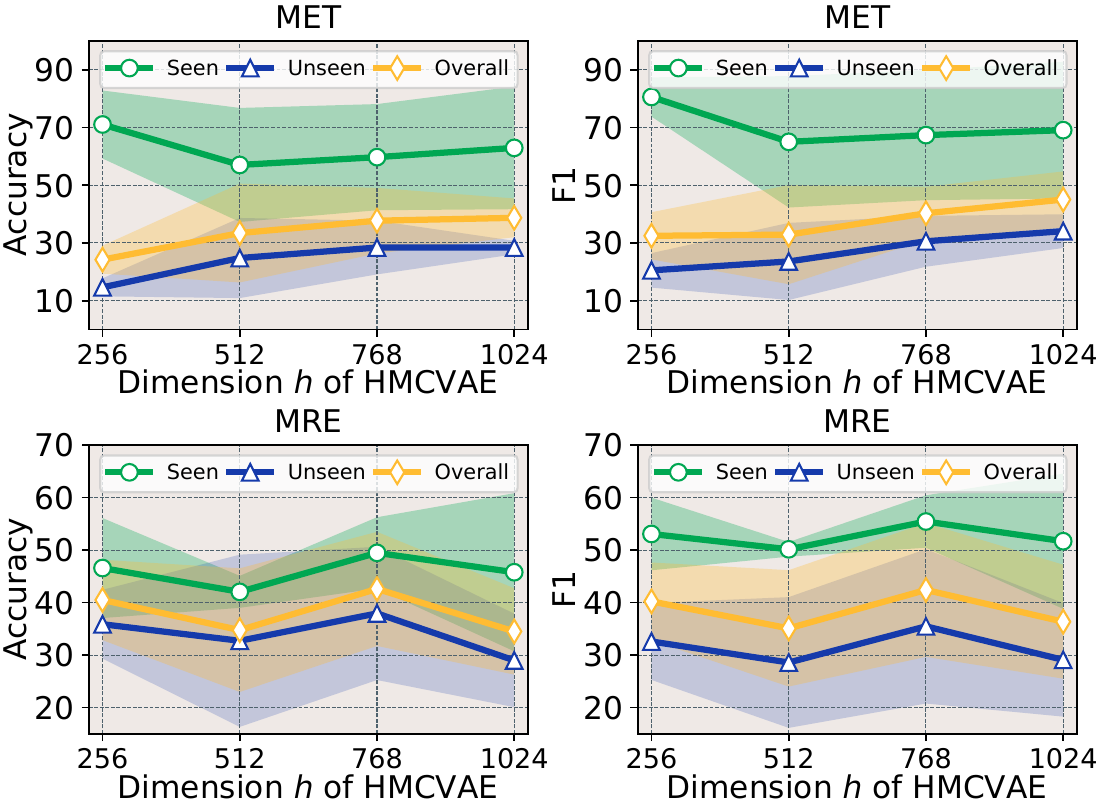}}
    \setlength{\abovecaptionskip}{0.2cm}
    \caption{Experimental results on the influence of Lorentz linear layer to the HMGRL framework.}
    \label{fig:ablation_hyperbolic_results}
    \vspace{-9pt}
\end{figure*}

\subsection{Ablation Study}
An ablation study was conducted to thoroughly evaluate the impact of the various modules incorporated in HMGRL, with the results presented in \tbref{tab:ablation-study}.
Notably, omitting the loss function for HMCVAE led to a significant drop in performance, highlighting the critical role of modeling multimodal representations of samples and generating synthetic samples for unseen categories.
To evaluate the influence of multimodal representation alignment, we individually removed HVIB and the loss term $\mathcal{L}_{cl}$, and observed that the models lacking these components all suffered from performance declines to varying extents. HVIB exerts a greater impact than the loss $\mathcal{L}_{cl}$ on the MET dataset, but the opposite is true on the MRE dataset.
Additionally, we assess the impact of exploiting the unseen categories with the losses $\mathcal{L}_{ce}$ and $\mathcal{L}_{align}$.
Our findings indicate that the utilization of the unseen categories plays a crucial role in boosting the model's final performance.
The corresponding results demonstrate that the classification on the synthetic samples of unseen categories is important to enhancing the model's final performance. Compared with the cross-entropy loss, the semantic distribution alignment loss contributes to a certain degree of improvement in the generalization capability of our proposed framework.
To summarize, the results of the ablation study validate that integrating HVIB and HMCVAE in conjunction with auxiliary losses for modeling multimodal representations of samples can improve model performance on GZS-MIE tasks.

\subsection{Influence of Hyper-parameters $\eta$ and $\zeta$}
In the proposed framework, the utilization of unseen categories encompasses both classification and semantic distribution alignment. To strike a balance between the utilization of real and synthetic samples, we introduce the hyper-parameters $\eta$ and $\zeta$.
To assess the impact of the hyper-parameter $\eta$, we perform parameter sensitive experiments as shown in \figref{fig:eta_results}.
The hyper-parameter $\eta$ governs the extent to which HMGRL emphasizes the classification of unseen categories. As the model allocates more focus to classifying unseen categories, the performance on seen categories may decline due to the varying quality of synthetic samples.
An excessively large $\eta$ value substantially impairs the overall performance of HMGRL on the MET dataset, as the model's capability for seen categories is adversely affected by its focus on unseen categories.
Besides, the hyper-parameter $\zeta$ influences the degree of semantic distribution alignment, we also perform parameter sensitive experiments to assess the impact of it as shown in \figref{fig:zeta_results}.
As the $\zeta$ value rises, the model's performance declines, and overemphasis on semantic distribution alignment results in a certain level of semantic matching failure, attributable to the continuous nature of the semantic representation space.

\subsection{Influence of Lorentz Linear Layer}
In order to evaluate the effect of the Lorentz linear layer on the proposed framework, the comprehensive experiments were conducted to analyze HVIB and HMCVAE.
Firstly, we reduced the Lorentz linear layers in HVIB and HMCVAE respectively and the ablation results are shown in \figref{fig:lll_ablation_study}.
The proposed framework is capable of attaining superior results compared to those achieved by models incorporating VIB or VAE, specifically in the ``Unseen'' and ``Overall'' items.
The aforementioned findings highlight the significant role the Lorentz linear layer plays in contributing to the final outcomes. Furthermore, given that the Lorentz linear layer is employed for encoding latent representations, we carried out experiments to investigate the influence of the dimensionality of these latent representations on HMGRL.
As shown in \figref{fig:lll_vib_dim}, the performance of HMGRL exhibits a trend where it initially improves and subsequently declines as the dimension of the latent representation in HVIB increases.
Besides, as illustrated in \figref{fig:lll_vae_dim}, the performance of HMGRL is notably impacted by the increase of the latent representation dimension in HMCVAE, as it directly influences to model multimodal representations and produce synthetic training samples.
The experimental results mentioned earlier provide additional evidence to support the role of the Lorentz linear layer in influencing generative representation learning within the framework we have proposed.

\section{Conclusion}
This study delves into generalized zero-shot multimodal information extraction (GZS-MIE) tasks, with the objective of tackling the challenges posed by hierarchical semantic representation and the gap in semantic similarity distribution. To overcome the disadvantages, we propose the hyperbolic multimodal generative representation learning (HMGRL) framework.
The HMGRL framework consists of the hyperbolic variational information bottleneck (HVIB) and hyperbolic multimodal conditional variational autoencoder (HMCVAE) with the auxiliary losses.
The HVIB network leverages Lorentz linear layers (L.L.L) to capture the hierarchical semantic interconnections among diverse modalities, facilitating the extraction of aligned multimodal representations.
The HMCVAE network can capture the hierarchical semantic correlation between samples and prototypes with L.L.L, thereby empowering it to model the representations of samples and generate synthetic instances of unseen categories. This facilitates training HMGRL with the semantic distribution alignment loss for enhanced performance.
Experimental results demonstrate the superiority of HMGRL on GZS-MIE tasks.

%%
%% The acknowledgments section is defined using the "acks" environment
%% (and NOT an unnumbered section). This ensures the proper
%% identification of the section in the article metadata, and the
%% consistent spelling of the heading.
\begin{acks}
This research was supported by the National Natural Science Foundation of China (No. 72342017) and the Fundamental Research Funds for the Central Universities, Nankai University (63253232).
\end{acks}

%%
%% The next two lines define the bibliography style to be used, and
%% the bibliography file.
\bibliographystyle{ACM-Reference-Format}
\bibliography{refs}

\appendix

\section{Appendix}

\subsection{Compared Methods and Evaluation Metrics} \label{appendix:compared_methods}
For ZS-IE, the prototype (Proto)~\cite{DBLP:conf/coling/MaCG16} network is a label embedding-based method to tackle the two tasks.
Moreover, the attention-based DBZFET~\cite{DBLP:conf/naacl/ObeidatFST19} and NZFET~\cite{DBLP:conf/www/RenLZ20} models, and the memory augmentation one MZET~\cite{DBLP:conf/coling/ZhangXLY20} are selected as the ZS-ET baselines.
Besides, the representation-based ZS-BERT~\cite{chen-li-2021-zs} model, 
the fine-grained semantic matching model RE-match~\cite{DBLP:conf/acl/ZhaoZZZGWWPS23}, and the prompt-based ZS-SKA~\cite{DBLP:conf/coling/GongE24a} model are considered as the ZS-RE baselines.
For ZS-MIE, we select the multimodal prototype network (MMProto)~\cite{DBLP:conf/aaai/WanZDHYP21}, the object visual context-based model MOVCNet~\cite{zhang-etal-2023-incorporating-object} and the fine-grained multimodal representation learning-based model MG-VMoE~\cite{DBLP:conf/www/ZhouZZSY25} as the multimodal baselines.
We also select 7B version of LLaVA (v1.5)~\cite{DBLP:conf/nips/LiuLWL23a} as the baseline to perform the GZS-MIE tasks.
For method evaluation, we use accuracy and weighted F1 score as metrics, with test set results reported separately for seen and unseen categories. The overall metrics are calculated by the harmonic means of accuracy and F1 scores on seen and unseen categories~\cite{DBLP:journals/pami/XianLSA19}.

\begin{figure}[t]
    \centering
    \includegraphics[width=\columnwidth]{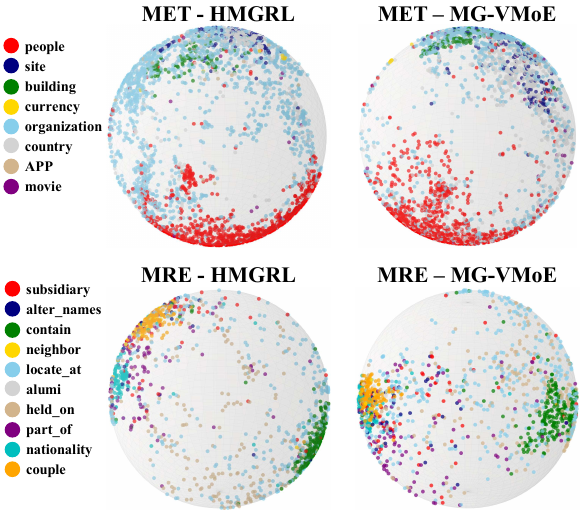}
    \caption{
        Visualization of multimodal features $\tilde{\textbf{e}}$ from GZS-MIE models via t-SNE projection onto a spherical manifold.
    }
    \label{fig:feature_analysis}
    \vspace{-9pt}
\end{figure}

\subsection{Visualization Analysis}
To assess the effectiveness of representation learning for GZS-MIE tasks, we visualize the representations $\tilde{\textbf{e}}$ extracted by the HMGRL and MG-VMoE models, as shown in \figref{fig:feature_analysis}. The samples from both seen and unseen categories of the test set were fed into the two models to obtain their respective multimodal representations. These features are then projected onto a spherical manifold using t-SNE~\cite{2008Visualizing} for dimensionality reduction.
The results demonstrate that HMGRL produces more cohesively clustered representations within specific categories than MG-VMoE. For example, on the MET dataset, multimodal representations for categories such as ``people'' and ``country'' exhibit tighter intra-category grouping. Additionally, when compared to MG-VMoE on the MRE dataset, HMGRL's representations show stronger category-level distinguishability.
Hence, the visualization analysis confirms the efficacy of HMGRL in modeling multimodal representations for GZS-MIE tasks.

\end{document}